\documentclass[preprint,12pt]{elsarticle}

\usepackage{amssymb}
\usepackage{amsmath}
\usepackage{nicefrac}
\usepackage[utf8]{inputenc}
\usepackage{amssymb}
\usepackage{nomencl}
\usepackage{xpatch}

\newlength{\nomitemorigsep}
\makenomenclature

\usepackage{etoolbox}

\renewcommand{\nomgroup}[1]{%
	\itemsep\nomitemorigsep%
	\ifthenelse{%
		\equal{#1}{A}%
	}{%
		\item[\textbf{Sets and Indices}]%
		
	}{%
		\ifthenelse{\equal{#1}{B}}{%
			\item[\textbf{Parameters and Constants}]%
			
		}{}%
	}{%
	\ifthenelse{\equal{#1}{C}}{%
		\item[\textbf{Variables}]%
		
	}{}%
}%
	\itemsep\nomitemsep
} 

\usepackage{setspace}
\usepackage{xcolor}
\usepackage{textcomp}
\usepackage{color,soul}

\allowdisplaybreaks
\journal{IJEPES}

\begin{document}

\begin{frontmatter}



\title{Optimal Participation of Price-maker Battery Energy Storage Systems in Energy and Ancillary Services Markets Considering Degradation Cost}


\author[inst1]{Reza Khalilisenobari}

\affiliation[inst1]{organization={School of Electrical, Computer and Energy Engineering, Arizona State University},
            city={Tempe},
            state={AZ},
            country={USA}}

\author[inst1]{Meng Wu}

\begin{abstract}
This paper proposes a bi-level optimization framework to investigate the optimal market operation strategies of price-maker battery energy storage systems (BESSs) in real-time energy, spinning reserve, and pay as performance regulation markets, with a special focus on understanding BESS's excessive regulation market participation observed by several system operators and the impact of battery degradation cost on BESS market activities.
An accurate battery degradation cost function is integrated into the BESS's strategic bidding model. An automatic generation control (AGC) signal dispatch model is proposed to deploy AGC signals in the bi-level framework. This enables thorough studies for BESS's operating characteristics in the frequency regulation market, when both battery degradation and detailed AGC signal following activities are considered. Case studies on a synthetic system with real-world data are performed to study interactions between BESS profit maximization and wholesale market operations, considering system annual load and ancillary service requirements variations. The impacts of BESS capacity and replacement cost on its revenue from energy and ancillary services markets are also investigated.
\end{abstract}




\end{frontmatter}

\printnomenclature[2.2cm]
\section{Introduction}

Motivated by US Federal Energy Regulatory Commission Order 841 \cite{Order841} requiring independent system operators (ISOs) to open energy and ancillary services markets for BESSs, many profit-seeking BESSs joined wholesale markets. From BESS market activities, several ISOs observe BESSs are very active in regulation markets, despite energy and reserve markets are also open \cite{ref31,IRENA2017}. In extreme cases, BESSs' regulation service almost covers system-wide regulation service requirements and saturates regulation markets \cite{news}. As BESSs can provide other services (energy, reserve, peak shaving, demand-side management, congestion management, etc.\cite{BESS_overview}) which, compared to regulation service, could be more important for stable/economic grid operations in some cases, it is desired to incentivize BESSs to provide multiple services in a more balanced way without concentrating exclusively on regulation service. This calls for efforts to explore mechanisms behind profit-seeking BESSs' excessive regulation market activities despite all energy/ancillary services markets are open. 


Existing literature for BESS market participation falls into two categories. The first category \cite{ref18,ref15,ref19,ref17,ref14,ref30,new_paper,he2020power} models the BESS as a price-taker. Operational characteristics of battery units, wide range of services that BESSs are capable of providing \cite{IRENA2017,BESS_overview}, and capacity of large-scale (utility-scale) BESSs enable BESSs to affect market outcomes and act as strategic units in real-time markets. Hence, the price-taker modeling approach does not capture the impacts of BESSs' strategic behavior on market clearing results and cannot obtain thorough understanding for market activities of profit-seeking large-scale BESSs.

The second category \cite{PriceQuota,MvsT1,ref22,ding2017optimal,ref25,toubeau2020data,ref24,MvsT2,MPQC} studies BESSs' strategic behavior by the price-maker BESS model, which is a bi-level optimization with interactions between BESSs' optimal bidding and ISO's wholesale market clearing. However, most works \cite{PriceQuota,MvsT1,ref22,ding2017optimal,ref25,toubeau2020data,ref24} overlook the regulation market and only model BESSs in the energy market \cite{PriceQuota,MvsT1,ref22,ding2017optimal,ref25} or energy and reserve markets \cite{toubeau2020data,ref24}. Since BESSs are very active in regulation markets \cite{ref31,IRENA2017}, these works ignore the BESSs' major revenue stream (the regulation market) and cannot accurately evaluate BESSs' strategic behavior and revenue streams in all three markets. Models in \cite{ref22,ding2017optimal,ref25,toubeau2020data,ref24,MvsT2} do not properly consider impacts of BESS degradation on their market activities, by either neglecting the degradation cost \cite{ref22,ding2017optimal} or modeling it by a linear function and/or cycling limitation model \cite{ref25,toubeau2020data,ref24,MvsT2} which represents BESS degradation in a less accurate way. Since BESSs have different degradation characteristics when providing different grid services (especially regulation services which requires less BESS cycling depth compared to other services), accurately modeling degradation costs could improve the understanding for BESSs' excessive regulation market participation. Models in \cite{MvsT2,MPQC} consider strategic BESSs in regulation markets (BESSs' major revenue stream). In \cite{MvsT2,MPQC}, AGC signals deployment is not considered in the regulation market models and regulation mileage deployment is modeled with predetermined factors. This modeling approach is not an accurate representation of the real-world regulation market structure and also does not capture the impact of following AGC signals on the battry degradation cost. Besides, Reference \cite{MvsT2} replaces the lower-level market clearing process in the bi-level optimization by predicting a `price quota curve' \cite{PriceQuota}. This price quota curve determines how the market clearing price (MCP) changes depending on the quantity offer of a strategic participant. Compared to the full bi-level model with the clearing process of energy/ancillary services markets, this price quota curve may not capture impacts of BESS's provision of one service (regulation) on the price of another service (reserve). Therefore, it cannot capture correlations between BESSs' strategic activities in different markets.  

To improve current market designs and incentivize BESSs toward providing multiple grid services in a more balanced and effective manner, a comprehensive modeling framework is needed to investigate the BESS's excessive regulation service provision despite all the energy, reserve, and regulation markets are available to BESSs. Several gaps remain between the existing literature and this desired framework. 1) The regulation market clearing details, such as the deployment of automatic generation control (AGC) signals, and the interactions between the regulation market and energy/reserve markets need to be modeled properly to capture BESS market activities; 2) The variations of BESS degradation costs when participating in different markets need to be explicitly considered; 3) The BESS's expected long-term market revenue, considering seasonal load variations, need to be investigated.

To resolve these gaps, this paper proposes a comprehensive framework for studying price-maker BESSs' strategic behavior in correlated energy, reserve, and regulation markets, with a special focus on investigating BESSs' excessive regulation service provision. This model is developed upon our previous work \cite{NAPS} with a simplified model (without BESS degradation cost and AGC signal dispatch model). Built upon existing bi-level optimization models for studying price-maker BESS's strategic behavior in energy/ancillary services markets \cite{PriceQuota,MvsT1,MPQC,toubeau2020data,ref25,ref22,ref24,MvsT2,ding2017optimal,NAPS}, this paper further makes the following contributions.
\begin{itemize}
	\item A realistic degradation cost model based on rainflow algorithm is deployed in the bi-level optimization framework to investigate BESS degradation cost variations when providing energy, reserve, and regulation services.  
	\item To study impacts of BESS's AGC signal following activities on its revenue, degradation, and operating patterns, an AGC signal dispatch model is proposed to deploy AGC signals in the bi-level framework based on market outcomes. This modeling effort enhances the regulation market clearing model and enables detailed investigations on BESSs' excessive regulation service provision.	
	\item BESS's operating characteristics across energy/ancillary services markets, BESS's regulation market participation strategies, the impact of BESS's profit maximization on wholesale market operations, and the impacts of BESS capacity, replacement cost and variations in load and reliability requirements of the grid on the BESS operation and revenue are analyzed thoroughly with this detailed modeling framework on a synthetic test case with real-world data for market parameters and BESS parameters.
\end{itemize}

The paper is organized as follows. Section 2 presents the bi-level model considering details of BESS AGC signal following activities and degradation. Section 3 presents simulation procedure and case studies. Section 4 concludes the paper.

\section{Formulation and Methodology}


This section formulates the bi-level framework for investigating the interactions between a strategic price-maker BESS's profit maximization and ISO's joint energy, reserve, and regulation markets clearing, considering 1) the degradation cost model deployed into the bi-level framework; 2) the detailed regulation market clearing process with the proposed AGC signal dispatch model. This framework contains the coupled upper-level problem (ULP) and lower-level problem (LLP). The ULP maximizes BESS's total profit across various markets. The LLP simulates ISO's joint market clearing process. Decision variables of the ULP, including BESS's quantity and price offers, serve as the input parameters for the LLP. The MCPs and BESS's scheduled power, determined by the LLP, serve as input parameters for the ULP. This bi-level problem is non-convex and is converted to a single-level mixed-integer linear programming (MILP) problem following the procedure in \cite{ref28}. This MILP can be solved by commercial solvers. Detailed ULP and LLP formulations are presented below after introducing the structure of the modeled markets.


\subsection{Market Structures}
This section describes the market structures for the real-time energy, spinning reserve, and pay as performance frequency regulation markets that are modeled in this stduy.

\subsubsection{Energy Market} 

The real-time energy market model allows BESSs and other market participants (such as conventional generators) to submit economic supply bids, containing quantity and price offers, to fulfill system net demand in each interval. In line with some ISO's (such as California ISO - CAISO) practice for demand bids in real-time market \cite[Section 30]{CAISOTarif}, it is assumed that demands have self-scheduling bids in real-time energy market, which means that they are willing to buy electricity regardless of the price. Hence, BESS's demand price offers must be high enough to ensure that their demanded quantity is cleared. In other words, BESS owner is a strategic participant while selling energy and is a price-taker participant, like other loads in the gird, when buying energy.

\subsubsection{Spinning Reserve Market}

According to the current definition of spinning reserve service in most ISOs (such as CAISO and PJM Interconnection) \cite[Section 8]{CAISOTarif}, \cite[Section 4]{M11PJM}, the modeled real-time spinning reserve market is an upward-only market. Although it is not usual to consider a downward spinning reserve market, it is similar to the upward reserve and can be handled by the proposed framework through minor adjustments. Market participants (including BESSs) are compensated by the MCPs for their reserved capacities. Deployment of spinning reserve product is not modeled in this study since the reserve deployment is called in the contingency situations, which falls out of the scope of this work. However, to guarantee the reliable operation of the system, the worst-case scenario of reserve deployment from BESS is considered in this study.

\subsubsection{Frequency Regulation Market}The pay as performance frequency regulation market consists of payment components for regulation capacity and regulation mileage. Capacity price is paid to regulating units for reserving each MW of their generation capability in order to enable regulation service provision. Mileage price compensates regulating units for every MW deployed toward following the AGC signal for both up- and down-regulation sent out by ISOs continuously (every four seconds). In ISO practice, the system-level AGC signal is dispatched to each regulation market participant, based on the regulation market outcome. The regulation market participants then follow these dispatched AGC signals (i.e., the unit-level AGC signals) for regulation mileage provision. Since it is a pay as performance market, each units' payments may be affected by its accuracy in following AGC signals \cite{ref27}.

In the regulation market model of this study, similar to the CAISO practice \cite{CAISOFRM}, the regulation mileage awarded to each regulating unit is lower- and upper-bounded respectively by the units' awarded regulation capacity and a factor of the regulation capacity (see Constraints (L5) and (L9)). This factor is called mileage multiplier, which is greater than one and is determined by ISO based on the historical performance and ramping capability of the regulation unit. Hence, while regulation units submit quantity and price offers for providing regulation capacity, they only submit price offers (without quantity offers) for regulation mileage provisions since the awarded regulation mileage is determined by the ISO based on units' regulation capacity, historical performance, and ramping capability.

Since a price-maker approach is adopted in this study, all the introduced markets are co-optimized in the LLP to model the interaction of BESS with markets and grid. In the price maker modeling approach, it is common to assume that the BESS owner can perfectly forecast grid requirements including AGC signals, and other participants' bids in markets \cite{MvsT1,ref22,ref25,ref24,MvsT2}. As the duration of real-time market clearing process varies from 5 to 15 minutes for different ISOs, the proposed framework has the flexibility of setting any duration for market clearing intervals. The duration of AGC signals dispatch within each market clearing interval is also adjustable in this framework.    

\subsection{The Upper-level Problem (ULP) Formulation}

In the ULP shown below, the BESS owner with several battery units at different buses maximizes its revenue from real-time energy, spinning reserve, and frequency regulation markets while considering its operating limits and degradation cost. 

\begin{align}
	&\text{Max} \sum_{t\in \mathcal{T}}\sum_{i \in \mathcal{B}}\Big\{\Big[\pi_{i,t}^{E}P_{i,t}^{B,E}+\pi_t^{Rs}P_{i,t}^{B,Rs}+\pi_{t}^{RgC}P_{i,t}^{B,RgC}+\pi_t^{RgM}P_{i,t}^{B,RgM} \Big] \Delta t\notag\\
	& \quad\quad - \sum_{z \in \mathcal{Z}}\sum_{k \in \mathcal{K}}C_{i,k}^{Deg}P_{i,t,z,k}^{Dis} \Delta z \Big\} \tag{U1}\\
	&\text{Subject to:}  \notag \\ 
	& {0 \leq Q_{i,t}^{SE} \leq u_{i,t}P_i^{Rate}}   \tag*{{$\forall i \in \mathcal{B}, \forall t \in \mathcal{T}$ (U2)}} \\
	& {0 \leq Q_{i,t}^{DE} \leq (1-u_{i,t})P_i^{Rate}}   \tag*{{$\forall i \in \mathcal{B}, \forall t \in \mathcal{T}$ (U3)}} \\
	& {\beta_{i,t}^{E} \geq (1-u_{i,t})M}   \tag*{{$\forall i \in \mathcal{B}, \forall t \in \mathcal{T}$ (U4)}} \\
	& 0 \leq Q_{i,t}^{Rs} \leq P_i^{Rate} \tag*{$\forall i \in \mathcal{B}, \forall t \in \mathcal{T}$ (U5)} \\
	& 0 \leq Q_{i,t}^{RgC} \leq P_i^{Rate} \tag*{$\forall i \in \mathcal{B}, \forall t \in \mathcal{T}$ (U6)} \\
	& -P_i^{Rate}+P_{i,t}^{B,RgC} \leq P_{i,t}^{B,E}\leq P_i^{Rate}-P_{i,t}^{B,RgC}-P_{i,t}^{B,Rs} \tag*{$\forall i \in \mathcal{B}, \forall t \in \mathcal{T}$ (U7)} \\
	& PF_{i,t}=\frac{P_{i,t}^{B,RgM}}{R_t^{RgM}} \tag*{$\forall i \in \mathcal{B}, \forall t \in \mathcal{T}$ (U8)} \\
	& \sum_{i \in \mathcal{B}}PF_{i,t} \leq 1 \tag*{$\forall t \in \mathcal{T}$ (U9)} \\
	& R_{{t}}^{RgM}=\sum_{{z}}|AGC_{{t},{z}}-AGC_{{t},{z}-1}| \tag*{$\forall t \in \mathcal{T}, \forall z \in \mathcal{Z}$ (U10)} \\
	& P_{i,t}^{B,E}+PF_{i,t}AGC_{t,z}=P_{i,t,z}^{TDis}-P_{i,t,z}^{TCh} \tag*{$\forall i \in \mathcal{B}, \forall t \in \mathcal{T}, \forall z \in \mathcal{Z}$ (U11)}\\
	& 0 \leq P_{i,t,z}^{TDis} \leq v_{i,t,z}P_i^{Rate} \tag*{$\forall i \in \mathcal{B}, \forall t \in \mathcal{T}, \forall z \in \mathcal{Z}$ (U12)}\\
	& 0 \leq P_{i,t,z}^{TCh} \leq (1-v_{i,t,z})P_i^{Rate} \tag*{$\forall i \in \mathcal{B}, \forall t \in \mathcal{T}, \forall z \in \mathcal{Z}$ (U13)}\\
	& P_{i,t,z}^{TDis}(\frac{1}{\eta_i})  = \sum_{k \in \mathcal{K}}P_{i,t,z,k}^{Dis} \tag*{$\forall i \in \mathcal{B}, \forall t \in \mathcal{T}, \forall z \in \mathcal{Z}$ (U14)}\\
	& P_{i,t,z}^{TCh}\eta_i = \sum_{k \in \mathcal{K}}P_{i,t,z,k}^{Ch} \tag*{$\forall i \in \mathcal{B}, \forall t \in \mathcal{T}, \forall z \in \mathcal{Z}$ (U15)}\\
	& P_{i,t,z,k}^{Ch}, P_{i,t,z,k}^{Dis} \geq 0 \tag*{$\forall i \in \mathcal{B}, \forall t \in \mathcal{T}, \forall z \in \mathcal{Z}, \forall k \in \mathcal{K}$ (U16)}\\
	& e_{i,t,z,k}-e_{i,t,z-1,k}=(P_{i,t,z,k}^{Ch}-P_{i,t,z,k}^{Dis})\Delta z \tag*{$\forall i \in \mathcal{B}, \forall t \in \mathcal{T}, \forall z \neq 1 \in \mathcal{Z}, \forall k \in \mathcal{K}$ (U17)}\\
	& e_{i,t,z,k}-e_{i,t-1,\bar{z},k}=(P_{i,t,z,k}^{Ch}-P_{i,t,z,k}^{Dis})\Delta z \tag*{$\forall i \in \mathcal{B}, \forall t \in \mathcal{T}, \tau =1, \forall k \in \mathcal{K}$ (U18)}\\
	& 0 \leq e_{i,t,z,k} \leq e_{i,k}^{Max} \tag*{$\forall i \in \mathcal{B}, \forall t \in \mathcal{T}, \forall z \in \mathcal{Z}, \forall k \in \mathcal{K}$ (U19)}\\
	& SOC_{i,t,z} = \sum_{k\in \mathcal{K}}e_{i,t,z,k} \tag*{$\forall i \in \mathcal{B}, \forall t \in \mathcal{T}, \forall z \in \mathcal{Z}$ (U20)}\\
	& SOC_i^{Min}+(P_{i,t}^{B,Rs}\Delta z) \leq SOC_{i,t,z} \leq SOC_i^{Max} \tag*{$\forall i \in \mathcal{B}, \forall t \in \mathcal{T}, \forall z \in \mathcal{Z}$ (U21)}\\
	& SOC_{i,t,z}-\sum_{t^*=0}^{t^*=t-1}P_{i,t^*}^{B,Rs}\Delta t - P_{i,t}^{B,Rs}z\Delta z \geq SOC^{Min} \tag*{$\forall i \in \mathcal{B}, \forall t \in \mathcal{T}, \forall z \in \mathcal{Z}$ (U22)}\\
	& SOC_{i,t,z}=SOC_i^{Init} \tag*{$\forall i \in \mathcal{B}, t = 0 \, \& \, z = \bar{z}, t = \bar{t} \, \& \, z = \bar{z}$ (U23)}\\
	& {u_{i,t}, v_{i,t,z}  \in \{0,1\}} \tag*{{$\forall i \in \mathcal{B}, \forall t \in \mathcal{T}, \forall z \in \mathcal{Z}$ (U24)}}
\end{align}

The objective function (U1) maximizes total revenue of a BESS owner (with multiple battery units at different buses). It considers the BESS's net revenue from the energy, reserve, regulation capacity and regulation mileage markets, as well as the BESS's degradation cost. While calculating the BESS's regulation capacity and mileage revenue, the BESS's performance score in following AGC signals is assumed to be 1, indicating the BESS's perfect AGC signals tracking accuracy due to its fast response capability. Decision variables of the ULP include each battery unit's price and quantity offers for real-time energy, spinning reserve capacity, regulation capacity, and regulation mileage provisions. These offers serve as inputs for ISO's joint market clearing process in the LLP. 

\subsubsection{Constraints for AGC Signal Dispatch Modeling}

In existing works, regulation mileage deployment (AGC dispatch) is usually modeled on the battery's state of charge (SOC) at the end of each market clearing interval (5 or 15 minutes) using predetermined factors of unit's regulation mileage participation $P_{i,t}^{RgM}$ \cite{MvsT2,MPQC,NAPS}. However, modeling SOC with this resolution (5 or 15 minutes) cannot capture SOC changes resulted from following AGC signals, since AGC signals are usually sent out every 4 seconds. 
It is important to monitor SOC changes resulted from following AGC signals to understand the effect of AGC following activities on BESS's degradation costs and operation. To accurately represent BESS's AGC signal following activities, the unit-level AGC signal dispatched to each battery needs to be properly modeled. Using (U8)-(U10), a participation factor (PF) is defined to simulate the AGC signal dispatch process and obtain the unit-level AGC signal for every battery. The proposed PF and the AGC signal dispatch simulation enable us to perform detailed studies on the impact of BESS's AGC signal following activities on its SOC management and battery degradation cost.  

In (U8), the PF is defined as the ratio of each battery's scheduled regulation mileage provision to system regulation mileage requirement. The summation of PFs for all batteries should not go beyond 1 (i.e., 100\% of the system regulation mileage requirement), as described in (U9). In (U10), the system regulation mileage requirement for a market clearing interval is defined by accumulating differences between two system-level AGC signal setpoints obtained at adjacent AGC sub-intervals. As mentioned, AGC signals are forecasted parameters of the optimization. Similar to the AGC dispatch method of \cite{ref26}, the concern of using PF for dispatching AGC signals is that the amount of dispatched AGC to each regulation unit does not exceed the unit's scheduled regulation capacity. This work addresses the issue by scaling the mileage multiplier $m$ of the LLP. These PFs are then used to calculate regulation mileage payments to various batteries (i.e., the BESS's revenue for regulation mileage provision). PFs are also used for calculating the charged/discharged power of each unit in each sub-interval by Constraint (U11).

\subsubsection{Constraints for Degradation Cost Modeling}

The battery degradation cost is typically calculated based on the rainflow algorithm, which counts the number of charge/discharge cycles, determines the cycle depth, and obtains accurate degradation cost based on the cycle count and cycle depth. 
However, this algorithm does not have an analytical expression and cannot be integrated directly in an optimization problem \cite{ref30}. This work deploys a linear approximation of the rainflow algorithm into the bi-level optimization framework. For degradation cost approximation, the battery capacity (from 0\% to 100\%) is uniformly divided into several segments (i.e., the degradation segments) represented by the $\mathcal{K}$ set. The energy capacity limit of each segment $e^{Max}$ is a portion of the battery capacity. A piecewise linear approximation of the degradation cost (obtained from the battery cycle depth aging function) is assigned to each segment $C^{Deg}$. This work calculates the degradation cost in each AGC signal sub-interval. Therefore, the degradation cost caused by following AGC signal and/or energy market participation is considered. Details on battery capacity segmentation, and $C^{Deg}$ calculation are in \cite{ref30}. Reference \cite{ref30} also shows that with sufficient number of segmentation, this linear approximation can be as accurate as exact rainflow algorithm.
In (U1), this degradation cost is subtracted from the BESS revenue. 

During a certain AGC signal sub-interval, Constraints (U11)-(U13) specify how much the battery unit is charging or discharging by determining the sign of the value obtained by summation of battery unit's scheduled power in the energy market and the AGC signal dispatched to the battery unit. In essence, Constraints (U11)-(U13) integrate the market clearing interval with AGC signal sub-interval by calculating the amount of charge/discharge power in each sub-interval, and this value is used in the remaining constraints for degradation cost calculation and operation scheduling of BESS. In each sub-interval, the battery unit may provide regulation up or down services while it sells or buys power in the energy market. The accumulation of these energy and regulation provisions determines the amount of power it is charging or discharging. For example, if in an interval, the battery unit buys 3 MW from energy market ($P_{i,t}^{B,E}=-3$) and receives 2 MW of AGC signal dispatch in one of the sub-intervals for regulation up service ($PF_{i,t}AGC_{t,z}=2$), then the left hand side of (U11) will be $-1$ resulting in $P_{i,t,z}^{TCh}=1$, which means that battery unit is charging for 1 MW during that sub-interval. Constraints (U14)-(U16) assign battery charge/discharge power during each AGC signal sub-interval to the corresponding degradation segments, while battery charge/discharge efficiency is considered. Constraints (U17)-(U18) evaluate battery stored energy in each degradation segment, by comparing difference between the stored energy in two consecutive AGC signal sub-intervals. Constraint (U17) performs this comparison for consecutive sub-intervals within the same market clearing interval, while Constraint (U18) performs this comparison between the first sub-interval of each market clearing interval and the last sub-interval of the previous market clearing interval. Constraint (U19) enforces the energy capacity limits of the degradation segments.

The above degradation cost representation (in (U1) and (U11)-(U19)) indicates in each AGC signal sub-interval, 1) if the battery unit is charging, the charged energy is allocated to the degradation segment with the lowest degradation cost which has not yet reached its maximum energy limit; 2) if the battery unit is discharging, the discharged energy is subtracted from the degradation segment with the lowest degradation cost which has been charged previously.

\subsubsection{Constraints for Battery SOC Management}

Constraint (U20) calculates the battery SOC during every AGC sub-interval by accumulating the stored energy in all segments of battery capacity. The energy in each segment is calculated through (U11)-(U19) based on the charged/discharged power resulted from following AGC signals and charged/discharged power in the energy market. Constraint (U21) limits the SOC value by the unit's maximum and minimum SOCs. In (U21), enough battery capacity is reserved for spinning reserve provision. Although spinning reserve deployment is not modeled, Constraint (U22) ensures that enough energy is stored in the battery to maintain reserve provision in the worst-case scenario for scheduled reserve deployment in all intervals. It guarantees the reliable operation of the system \cite{Worst_Res}. Constraint (U23) ensures the SOCs at the beginning and end of the simulation horizon remain at the same pre-specified value, so it is possible to expand a short-time simulation result to a longer period. Constraint (U24) defines binary variables for the battery's charging/discharging states during different sub-intervals $v_{i,t,z}$ and binary variables for the battery's supply/demand states during different market clearing intervals $u_{i,t}$. 

\subsubsection{Constraints for Battery Charge/Discharge Limits}
Constraints (U2)-(U3) limit unit $i$ of BESS to either sell ($u_{i,t}=1$) or buy energy ($u_{i,t}=0$) in the energy market, while its quantity offer is limited by the unit's charge/discharge rate. If it buys energy, Constraint (U4) ensures the battery's energy price offer $\beta_{i,t}^E$ is large enough such that the battery behaves as a self-scheduling unit like other loads in the system as described in Section 2.1. Constraints (U5)-(U6) limit the battery's quantity offers for spinning reserve and regulation capacity provisions within the unit's charge/discharge rate, respectively. Constraint (U7) indicates the battery's total power scheduled for energy, reserve capacity, and regulation capacity provisions must stay below the unit's charge/discharge rate. Note that regulation capacity provision is symmetrical and represents both regulation up and down.

\subsection{The Lower-level Problem (LLP) Formulation}
The LLP below models ISO's joint market-clearing process for real-time energy, spinning reserve, and pay-as-performance frequency regulation markets (with regulation capacity and mileage payments). 
\begin{align}
	&\text{Min} \sum_{t\in \mathcal{T}}\Big[\sum_{j\in \mathcal{G}} \big( \alpha_{j,t}^{{E}}P_{j,t}^{G,S}+\alpha_{j,t}^{Rs}P_{j,t}^{G,Rs}+\alpha_{j,t}^{RgC}P_{j,t}^{G,RgC}+\alpha_{j,t}^{RgM}P_{j,t}^{G,RgM} \big) +\notag\\
	& \qquad \: \qquad \sum_{i\in \mathcal{B}} \big(\beta_{i,t}^{E}P_{i,t}^{B,E}+\beta_{i,t}^{Rs}P_{i,t}^{B,Rs}+ \beta_{i,t}^{RgC}P_{i,t}^{B,RgC}+\beta_{i,t}^{RgM}P_{i,t}^{B,RgM}\big)\Big] \Delta t \tag{L1}\\
	&\text{Subject to:}  \notag \\ 
	& P_j^{Min}+P_{j,t}^{G,RgC} \leq P_{j,t}^{G,S} \leq  P_j^{Max}-P_{j,t}^{G,Rs}-P_{j,t}^{G,RgC} \tag*{$\forall j \in \mathcal{G}, \forall t \in \mathcal{T}$ (L2)} \\
	& 0 \leq P_{j,t}^{G,Rs} \leq P^{Rs,ramp}\tag*{$\forall j \in \mathcal{G}, \forall t \in \mathcal{T}$ (L3)}\\
	& 0 \leq P_{j,t}^{G,RgC} \leq P^{Rg,ramp}\tag*{$\forall j \in \mathcal{G}, \forall t \in \mathcal{T}$ (L4)}\\
	& P_{j,t}^{G,RgC} \leq P_{j,t}^{G,RgM} \leq m_{j,t}P_{j,t}^{G,RgC} \tag*{$\forall j \in \mathcal{G}, \forall t \in \mathcal{T}$ (L5)}\\
	& {-Q_{i,t}^{DE} \leq P_{i,t}^{B,E} \leq Q_{i,t}^{SE}}\tag*{{$\forall i \in \mathcal{B}, \forall t \in \mathcal{T}$ (L6)}}\\
	& 0 \leq P_{i,t}^{B,Rs} \leq Q_{i,t}^{Rs}\tag*{$\forall i \in \mathcal{B}, \forall t \in \mathcal{T}$ (L7)}\\
	& 0 \leq P_{i,t}^{B,RgC} \leq Q_{i,t}^{RgC}\tag*{$\forall i \in \mathcal{B}, \forall t \in \mathcal{T}$ (L8)}\\
	& P_{i,t}^{B,RgC} \leq P_{i,t}^{B,RgM} \leq m_{i,t}P_{i,t}^{B,RgC} \tag*{$\forall i \in \mathcal{B}, \forall t \in \mathcal{T}$ (L9)}\\
	& \sum_{j\in \mathcal{G}}P_{j,t}^{G,Rs} + \sum_{i\in \mathcal{B}}P_{i,t}^{B,Rs} \geq R_t^{Rs}: \pi_t^{Rs}\tag*{$\forall t \in \mathcal{T}$ (L10)}\\
	& \sum_{j\in \mathcal{G}}P_{j,t}^{G,RgC} + \sum_{i\in \mathcal{B}}P_{i,t}^{B,RgC} \geq R_t^{RgC}: \pi_t^{RgC}\tag*{$\forall t \in \mathcal{T}$ (L11)}\\
	& \sum_{j\in \mathcal{G}}P_{j,t}^{G,RgM} + \sum_{i\in \mathcal{B}}P_{i,t}^{B,RgM} \geq R_t^{RgM}: \pi_t^{RgM}\tag*{$\forall t \in \mathcal{T}$ (L12)}\\
	& \sum_{j\in \mathcal{G}|j=n}P_{j,t}^{G,S} + \sum_{i\in \mathcal{B}|i=n}P_{i,t}^{B,E}= P_{n,t}^{Load}+\sum_{w\in\mathcal{N}}H_{nw}(\theta_{n,t}-\theta_{w,t}): \pi_{n,t}^{E}\tag*{$\forall n \in \mathcal{N}, \forall t \in \mathcal{T}$ (L13)}\\
	& -TL_{nw} \leq H_{nw}(\theta_{n,t}-\theta_{w,t}) \leq TL_{nw} \tag*{$\forall (n,w) \in \mathcal{L}, \forall t \in \mathcal{T}$ (L14)}
\end{align}

The objective function (L1) determines the system's total operating cost. The decision variables are the scheduled power of each generating/battery unit for energy, reserve, regulation capacity, and regulation mileage provisions. The LLP also determines the energy, reserve, regulation capacity, and regulation mileage prices for each battery unit. The battery's scheduled power and MCPs serve as inputs for the ULP.

\subsubsection{Constraints for AGC Signal Dispatch Modeling}

Constraints (L5) and (L9) apply to each regulation market participant (BESSs and generators) to represent the pay as performance market model. The regulation mileage multiplier $m$ is assigned by the ISO to each regulating unit based on its historical performance on regulation service provision and its ramping capability. These multipliers serve as parameters of the optimization problem.

To ensure the AGC signal dispatched to each regulating unit (based on the participation factor $PF$) does not go beyond the unit's scheduled power for regulation capacity provision, we propose to uniformly scale the unit's mileage multiplier $m$ in a way that (1) always holds. Note that in most market clearing intervals, Equation (1) may hold without scaling.    
\begin{align}
	\label{m_scale}
	m_{n,t} \leq \frac{R_{t}^{RgM}}{Max_z(|AGC_{t,z}-AGC_{t,z-1}|)} \notag \\
	\tag*{$\forall n \in \mathcal{B} \cup \mathcal{G}, \forall t \in \mathcal{T}$ (1)}
\end{align}
\indent With (1), in each market clearing interval, the regulation mileage multiplier of each regulating unit is scaled to be less than the ratio of system regulation mileage requirement to the maximum change between consecutive AGC signal setpoints. Consider a system whose regulation capacity and mileage requirements are $R_t^{RgC}$ = 6MW and $R_t^{RgM}$ = 12MW, respectively, in an interval. Assume Units 1, 2, and 3 provide regulation service. They have identical 2MW scheduled regulation capacity ($P_1^{RgC}$ = $P_2^{RgC}$ = $P_3^{RgC}$ = 2MW). Based on historical performance, ISO sets their mileage multipliers as $m_1$ = 5, $m_2$ = $m_3$ = 3. Without uniform scaling of mileage multipliers, if Unit 1 has the lowest regulation mileage price bid and the other two units have identical bids, the units' scheduled regulation mileage will be $P_1^{RgM}$ = 8MW, $P_2^{RgM}$ = $P_3^{RgM}$ = 2MW (according to Constraint L5 or L9, lower limits for $P_1^{RgM}$ and $P_2^{RgM}$ are their regulation capacities, which are 2 MW). Based on (U8), the units' PFs are $PF_1=\frac{2}{3}, PF_2=PF_3=\frac{1}{6}$. If maximum change between consecutive system-wide AGC signals is 4 MW ($Max_z(|AGC_{t,z}-AGC_{t,z-1}|)$ = 4MW), Unit 1 should provide $\frac{2}{3}$ of it (2.6 MW), which is greater than the unit's scheduled regulation capacity and is not acceptable. By uniformly scaling mileage multipliers with a factor of $\frac{3}{5}$, the inequality in (1) is guaranteed, resulting in $m_1$ = 3, $m_2$ = $m_3$ = 1.8 (one can see that these new mileage multipliers are still prioritizing unit 1 over others with the same ratio). Accordingly, the units' scheduled regulation mileage will be $P_1^{RgM}$ = 6MW, $P_2^{RgM}$ = $P_3^{RgM}$ = 3MW. Hence, Unit 1 should provide 2MW of the 4MW system-wide AGC signal change, which lies within its scheduled regulation capacity. Note that Unit 1 can increase its share for regulation mileage provision by increasing its share in the regulation capacity market. As all mileage multipliers are uniformly scaled, their role, which is prioritizing units for regulation mileage provision, is not affected. The proposed uniform scaling of mileage multipliers is a representation of out of market adjustments that may happen in reality for ensuring that the dispatched AGC signals are less than each unit's regulation capacity. 

\subsubsection{Constraints for Market Participants}

Constraints (L2)-(L4) ensure for each generating unit, 1) its total power delivery lies within its maximum and minimum generation limits; 2) its reserve and regulation capacity provisions do not exceed the corresponding ramp rates. Constraints (L6)-(L8) ensure the battery's scheduled power in the energy, reserve, and regulation capacity markets lie within the corresponding quantity offers of the BESS (determined by the ULP).

\subsubsection{Constraints for Market Operations}

Constraints (L10)-(L12) ensure system requirements on reserve, regulation capacity, and regulation mileage services are satisfied. The dual variables of these constraints $\pi_t^{Rs}, \pi_t^{RgC}, \pi_t^{RgM}$ represent the MCPs for the corresponding services. Constraint (L13) represents the nodal power balance. Its dual variable $\pi_{n,t}^E$ represents the locational marginal price (LMP) at bus $n$. The above MCPs and LMPs serve as inputs for the ULP. Constraint (L14) represents the transmission line thermal limits.

\section{Case Studies}

\begin{figure}[!h]
	\centering
	\includegraphics[width=3.4in]{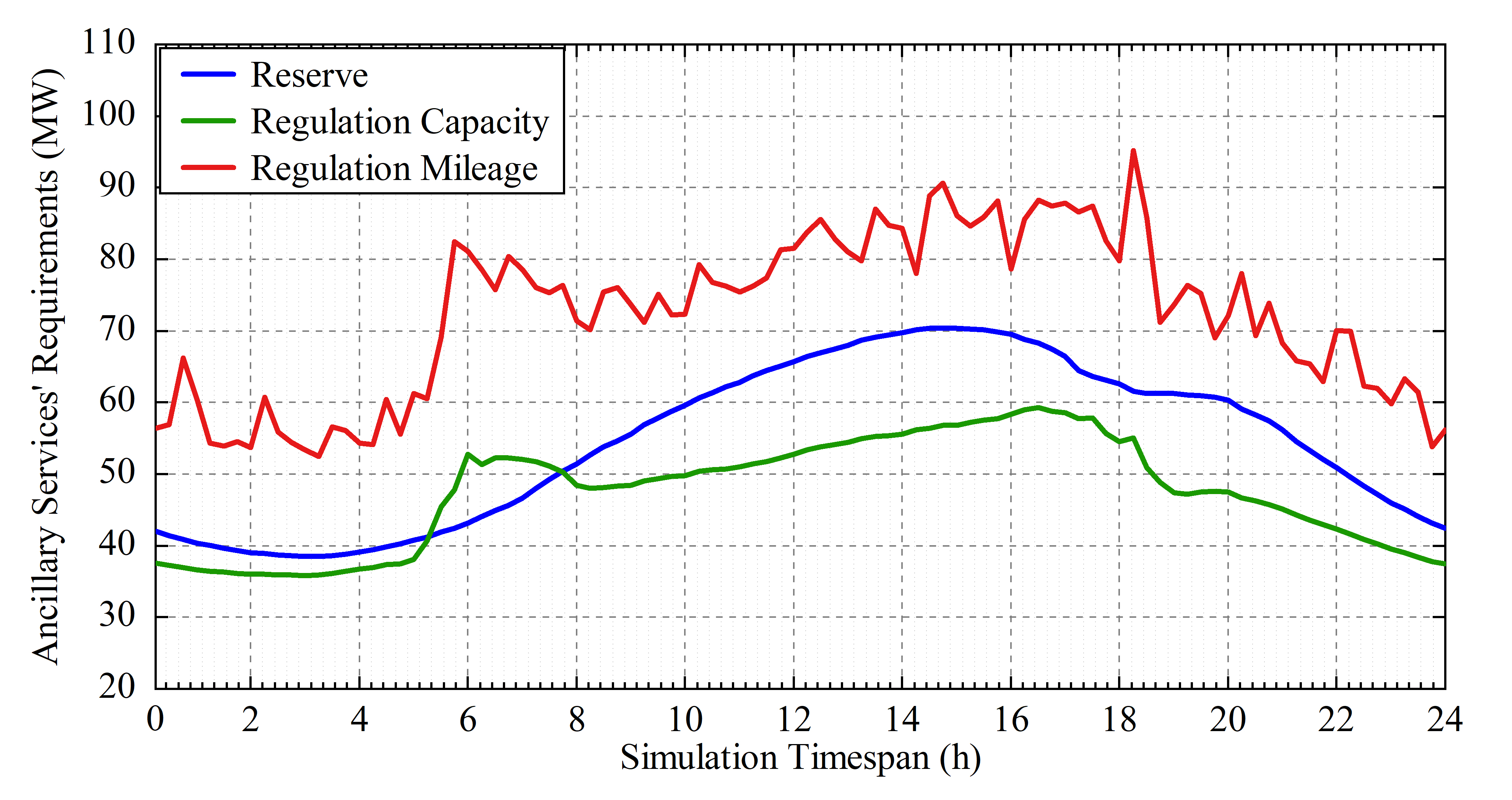}
	\caption{The test system's ancillary services requirements.}
	\label{ASR}
\end{figure}

The framework is evaluated on 3rd area of IEEE Reliability Test System of the Grid Modernization Laboratory Consortium (RTS-GMLC) \cite{RTS}, with 25 buses, 39 lines, 26 generators, and a BESS. The simulation horizon is 24 hours. The market clearing interval is 15 minutes. The AGC signal dispatch sub-interval is set at 20 seconds to reduce computational burden. System load profile and reserve/regulation capacity requirements are created using average load of the test case from June to August. System regulation mileage requirement in each market clearing interval is set at around 1.5 times corresponding regulation capacity requirement (a typical requirement in real-world markets \cite{ref26}). 
Fig. \ref{ASR} shows 24-hour system ancillary services requirements.
Sample AGC signals from ISO New England (ISO-NE) are adopted \cite{AGC} and modified to match regulation mileage requirement in each interval.

We assume generators use generation costs as energy price offers. Generator price offers for reserve, regulation capacity and mileage are created using their energy price offers multiplied by 0.15, 0.4, and 0.07, respectively, where the multipliers are corresponding average ratios in PJM historical data\cite{PJM}.


There is a lithium-ion BESS unit at Bus 13. Its operational parameters are shown in Table \ref{T1}. This battery has a 4-hour operation duration and an 80\% maximum charge/discharge cycle depth. This type of battery's useful life is 6000 cycles at 80\% charge/discharge cycle depth \cite{life}, and their replacement cost is around 200 k\$/MWh \cite{price}.



%
%

\begin{table}[!h]
	\caption{Operational Information of BESS's Unit}
	\label{T1}
	\centering
	\begin{tabular}{cccccc}
		\hline
		$P^{Rate}$ & Capacity & $SOC^{Min}$ & $SOC^{Max}$ & $SOC^{Init}$ & $\eta$\\
		(MW) & (MWh) & (MWh) & (MWh) & (MWh) & $\%$  \\
		\hline
		50 & 200 & 20 & 180 & 90 & 95\\
		\hline
	\end{tabular}
\end{table}

Based on lithium-ion BESS's cycle depth aging function \cite{deg-model} and replacement cost, the degradation cost function is $C(\delta)=52.4\,\delta^{2.03}$, where $\delta$ is the ratio of cycle depth to BESS capacity. This near quadratic function is approximated by a 16-segment piecewise linear function for degradation cost modeling. 

\subsection{BESS Degradation Cost and Regulation Market Activities}

\begin{figure}[!h]
	\centering
	\includegraphics[width=3.4in]{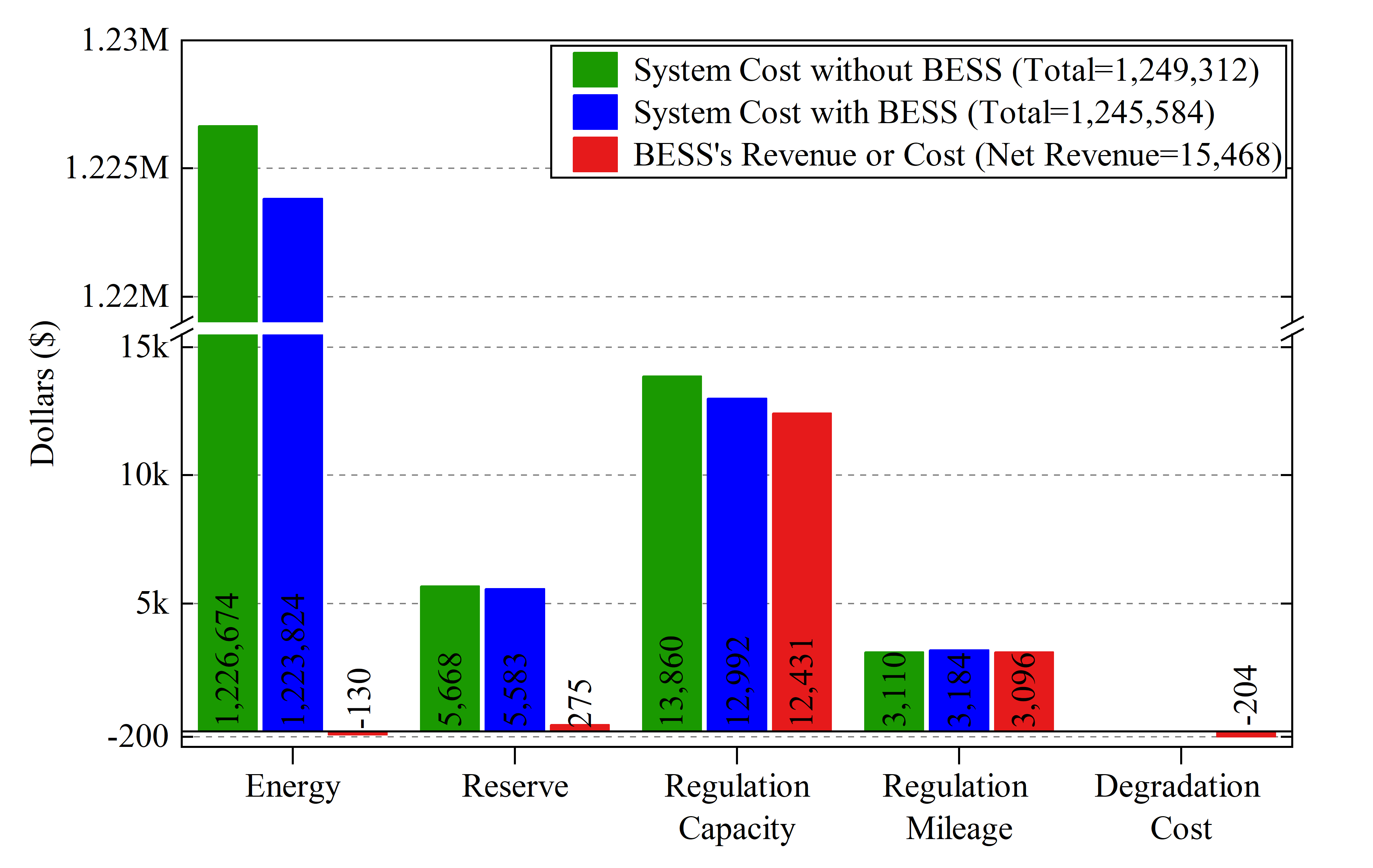}
	\caption{System cost with/without BESS and BESS's revenue and costs.}
	\label{rev}
\end{figure}

Fig. \ref{rev} shows 1) BESS total revenue from each market and its degradation cost; 2) system total operation cost for each market, with and without the BESS. 

It is observed in Fig. \ref{rev} that although a price-maker BESS seeks to maximize its profit, its market participation still reduces the system total operation cost and benefits the whole system. 
Comparing the total operation cost of the regulation capacity market and BESS's revenue from this market, it is clear that BESS is the major contributor to the regulation capacity market. However, with the BESS, the total operation cost of the regulation capacity market still reduces, and the market power of BESS is limited in this synthetic test case system.  
Fig. \ref{rev} shows almost all the BESS's revenue comes from the regulation capacity and mileage markets, indicating regulation markets being the most profitable for BESS. High share of BESS in regulation market agrees with real-world observations \cite{ref31,IRENA2017,news}. Besides, in Fig. \ref{rev}, BESS gains the least revenue from energy market. This is because energy arbitrage activity of BESS in energy market results in deep charge/discharge cycles, which is associated with high degradation cost. This reduces the BESS's revenue from the energy market. Note that although BESS does not have a considerable share in energy market, its participation in ancillary services markets re-allocates cheaper units to energy market and reduces the system operation cost of energy market. 



\begin{figure}[!h]
	\centering
	\includegraphics[width=3.4in]{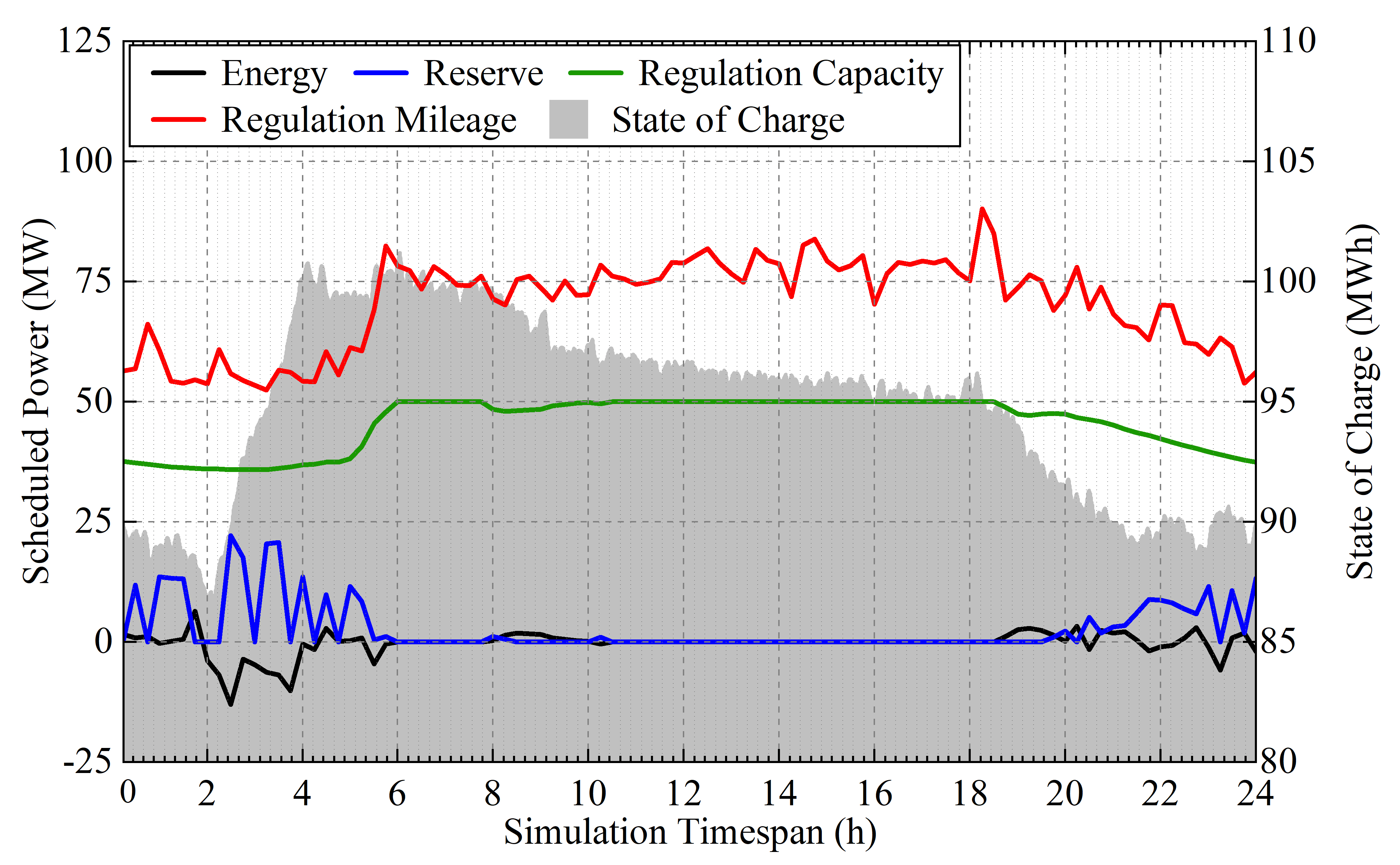}
	\caption{Scheduled power of BESS in each market and its state of charge.}
	\label{base}
\end{figure}

Fig. \ref{base} shows BESS scheduled power in each market and its SOC across 24 hours simulation horizon (96 market clearing intervals). 

The comparison between Fig. \ref{ASR} and Fig. \ref{base} shows that when system regulation capacity requirements are less than BESS charge/discharge limit (50MW), system-level regulation requirements are fully satisfied by BESS's regulation service. This indicates, compared to the energy and reserve markets, the regulation market is the primary option for the BESS. 
During hours 6-8 and 10-18, the system regulation capacity requirements exceed BESS charge/discharge limit, and the BESS could not provide all the required regulation services. In this case, the BESS scheduled power for regulation capacity provision is mostly fixed at the charge/discharge limit, and the system regulation requirements are jointly satisfied by the BESS and other market participants. 
At the beginning/end of the day when the system regulation capacity requirement falls below the charging/discharging limit, the BESS first provides all the system's regulation requirement, then optimally allocates its unused output capacity across the other markets. During these intervals, BESS performs energy arbitrage across various hours by charging itself in initial hours, which have low energy prices, to avoid buying energy (charging) during hours with high energy prices for maintaining the required end of the day SOC. Also, the BESS performs same interval energy arbitrage across energy and reserve markets during intervals when it is charging (buying energy) from grid by selling all or portion of its charging power as reserve capacity. 


The SOC curve in Fig. \ref{base} shows the BESS does not experience deep charge/discharge cycles. Extracting the charge/discharge cycles of this SOC curve using exact rainflow algorithm \cite{RainFlow} reveals the BESS goes through 207 charge/discharge cycles during the simulation. The cycle depth varies between 1.72MWh and 4kWh, and the average cycle depth is 0.23MWh. These charge/discharge cycles are very shallow (less than 1\% of the BESS capacity).
These shallow charge/discharge cycles are caused by the BESS's AGC signal following activities. As reducing cycle depths lowers the BESS degradation cost, participating in the regulation market could reduce the BESS's degradation cost while maintaining a certain revenue level. This agrees with Fig. \ref{rev}, in which the BESS degradation cost remains low.
Additionally, shallow charge/discharge cycles indicate the BESS does not use most of its capacity. Therefore, the BESS revenue depends more on its charge/discharge limit rather than its capacity.

\subsection{Variations of System Load and Reliability Requirements}

\begin{figure}[!h]
	\centering
	\includegraphics[width=3.4in]{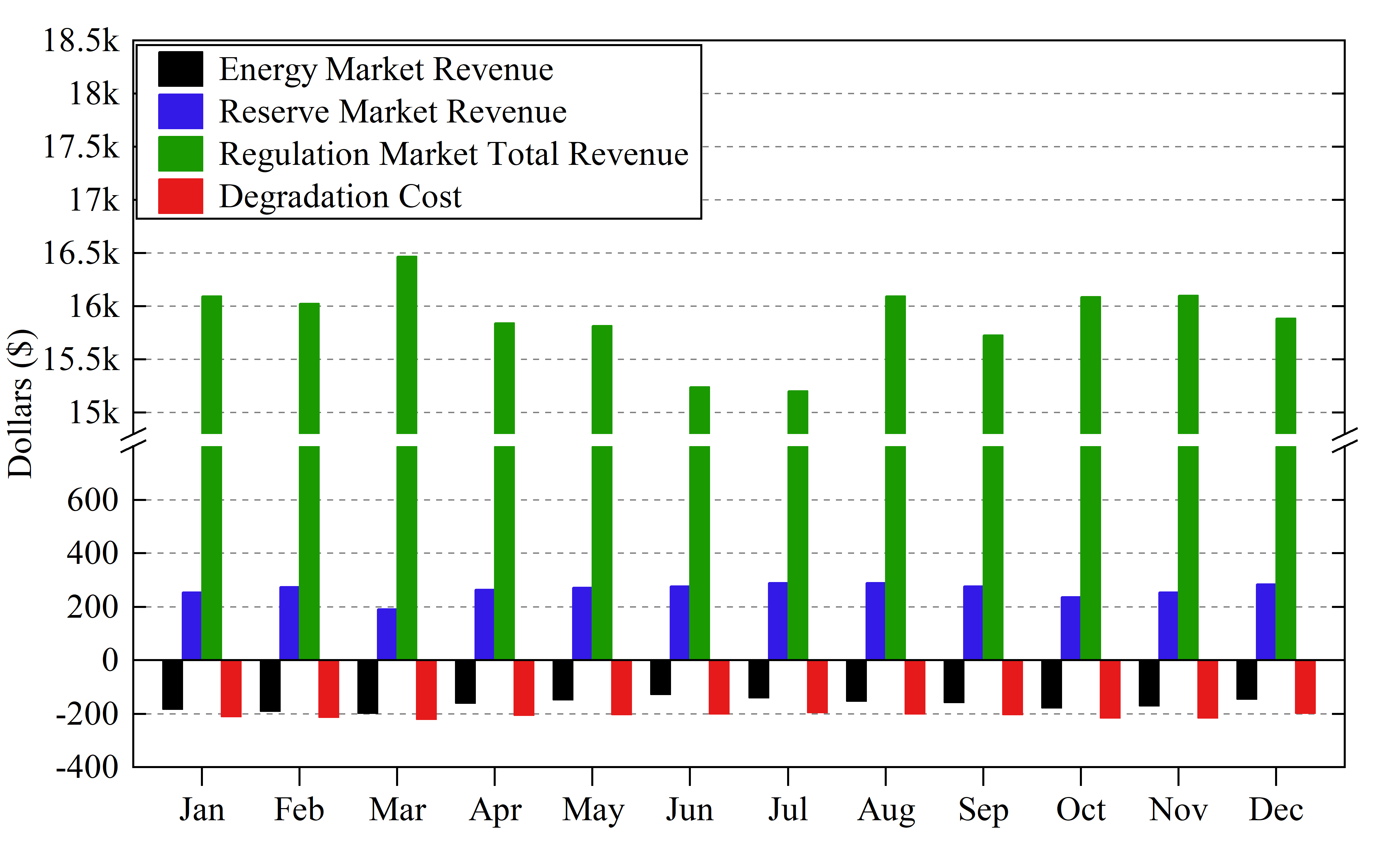}
	\caption{Degradation cost and revenue of BESS from each market in various simulations covering average monthly variations in load and reliability requirements of the system.} 
	\label{load}
\end{figure}

To study impacts of grid load/ancillary service requirements variations and AGC signal changes on BESS seasonal operations and long-term revenue, 12 simulations are performed in which 24-hour grid load/ancillary service requirements are the average of the test case load/ancillary service requirements in each month of the year. Across 12 simulations, the average/maximum/minimum changes of grid load in each interval are 360MW, 713MW, and 203MW, respectively, covering wide range of grid load/ancillary service requirements variations to fully analyze the sensitivity of BESS's operation to these parameters. Forecasted AGC signals are generated for each simulation based on regulation mileage requirement and synthetic AGC signals' pattern provided by ISO-NE \cite{AGC}. Hence, 288 different AGC hourly time series are used in these simulations to study the effect of AGC signal variations on BESS's operation and degradation cost. Fig. \ref{load} shows BESS revenue from each market and its degradation cost in 12 simulations.



According to Fig. \ref{load}, over various loads/ancillary service requirements, 1) BESS participates in the regulation market the most and the energy market the least; 2) its regulation market revenue is significantly higher than its reserve market revenue, while its energy market revenue is negative; 3) it participates in energy market mainly for buying energy to compensate its discharged energy; 4) the revenue pattern and market participation priority of BESS are consistent with variations in load/ancillary service requirements and AGC signals, which further verify the outcomes of previous section. With these simulations built upon averaged system load and ancillary service requirements for each month, the BESS's expected annual revenue (including regulation/reserve market revenue), annual costs (including degradation/energy cost), and annual net profit are estimated as \$5,910,000, \$135,000, and \$5,775,000, respectively.



In Fig. \ref{load}, BESS regulation market revenue varies the most compared to its revenue from other markets,
since BESS is a major contributor to regulation market. Its regulation market revenue is more affected by system regulation requirement variations.
BESS energy/reserve market revenue is less affected by system load/reserve requirement variations, due to its limited energy/reserve market participation.
During June/July when system regulation requirement is lower, BESS regulation market revenue is lower, while its energy/reserve market revenue is higher and degradation cost is lower. During this period, 1) BESS has extra capacity for reserve market participation; 2) it buys less energy to compensate energy discharged due to AGC signal following; and 3) it has lower degradation cost due to reduced AGC signal following.


\subsection{Analysis of the BESS's Capacity and Degradation Cost}

\begin{figure}[!h]
	\centering
	\includegraphics[width=3.4in]{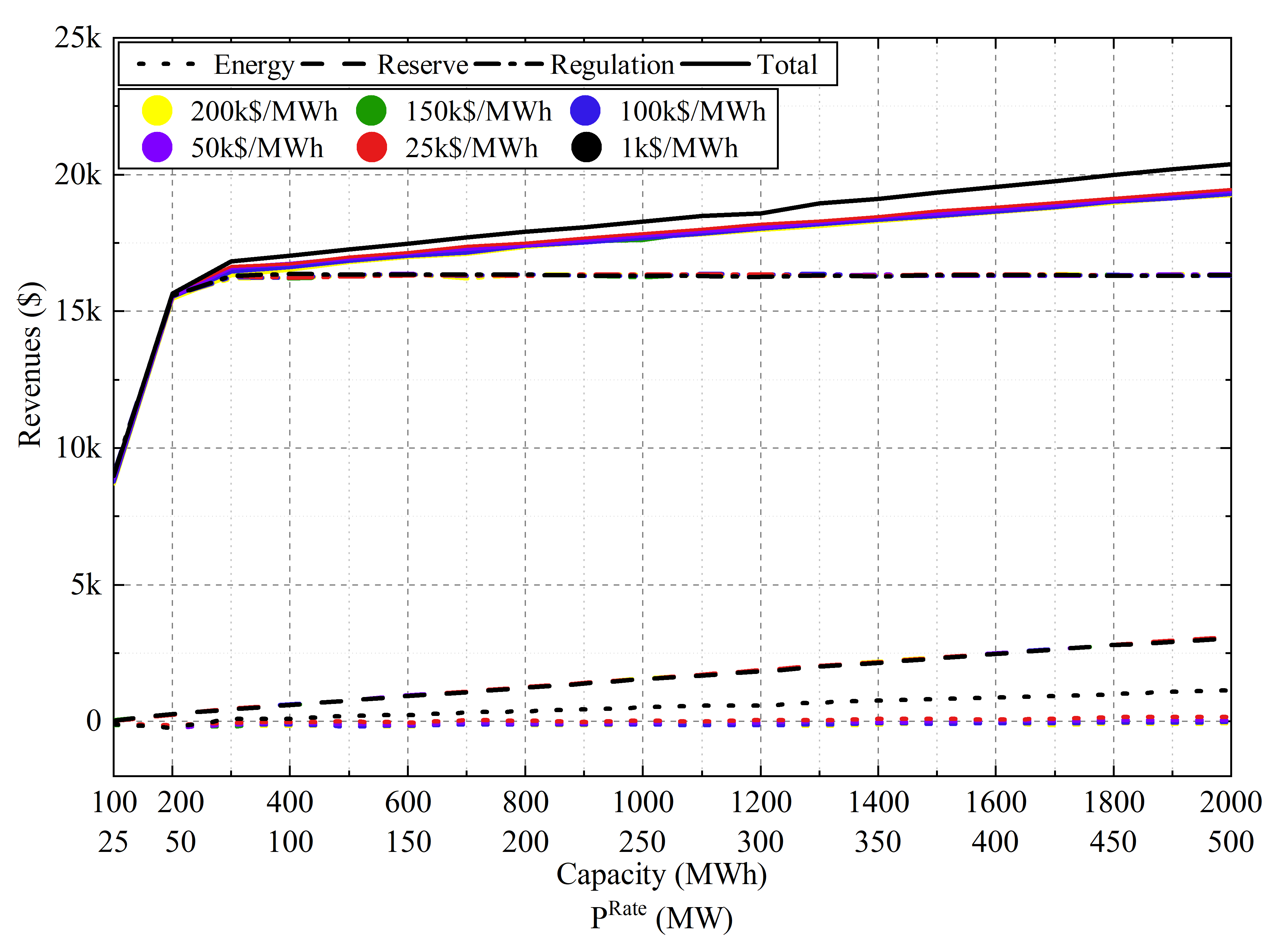}
	\caption{BESS's revenue versus its capacity and charge/discharge limit for different replacement costs. The BESS's total revenue and revenue from each market are represented using curves with specific patterns, and each curve is shown in different colors representing various battery replacement costs. E.g., the dotted black curve is the energy market revenue for a BESS with 1k\$/MWh replacement cost. The horizontal axis is labeled with both BESS capacity (in MWh) and charge/discharge limit (in MW).}
	\label{S&C}
\end{figure}

A bi-parametric analysis is performed to study impacts of BESS capacity and replacement cost on BESS operations and market revenue. 
BESS capacity is changed from 100MWh to 2000MWh (step size = 100MWh). With a 4-hour operation duration, BESS charge/discharge limit changes from 25MW to 500MW (step size = 25MW). For each size level, we consider replacement costs of 200k\$/MWh, 150k\$/MWh, 100k\$/MWh, 50k\$/MWh, 25k\$/MWh, and 1k\$/MWh. Degradation costs vary with the replacement costs. To maintain degradation cost modeling accuracy, the number of degradation cost segments is varied based on BESS capacity variations. We perform 120 simulations to cover these variations. For each replacement cost, BESS total revenue and revenue from each market versus its capacity (and charge/discharge limit) are shown in Fig. \ref{S&C}. 

In Fig. \ref{S&C}, regulation market revenue curves at various replacement costs increase as BESS capacity increases to 300MWh. Beyond 300MWh capacity, system regulation requirements are fully satisfied (i.e., saturated) by BESS. Hence, BESS regulation market revenue curves remain flat after 300MWh capacity. A knee-point happens on BESS total revenue curves when regulation market gets saturated. This indicates BESS total revenue increases slower when it cannot gain more profit from the regulation market. In existing systems, after expanding BESS capacity beyond a certain threshold, BESS total revenue growth rate may decrease significantly as the BESSs saturate the regulation market and cannot gain significant revenue from the energy/reserve markets.



In Fig. \ref{S&C}, except for total revenue curve and energy revenue curve at 1k\$/MWh replacement cost, all other curves of the same pattern at different replacement costs overlap. This indicates 1) BESS replacement cost and its revenue from reserve/regulation markets are not strongly correlated;
2) BESS energy market participation is not profitable unless the replacement cost drops significantly (to 1k\$/MWh). Once BESS energy market participation becomes profitable, after saturating the regulation market, BESS allocates its capacity in excess of ancillary services provision to the energy market.


In Fig. \ref{S&C}, regardless of the replacement cost, BESS reserve revenue increases almost constantly as its capacity increases, since BESS reserve market participation is bounded by its capacity due to worst-case reserve deployment model. Its regulation revenue (before saturation) increases significantly faster than its energy/reserve revenue. This indicates BESS capacity and charge/discharge limit increment has more impact on its regulation revenue than on its energy/reserve revenue.

These results show: 1) compared to energy market, regulation/reserve markets are more profitable and have higher participation priority for BESSs, regardless of replacement costs; 2) participation of BESS in energy market is not profitable unless battery replacement cost drops significantly, which is neglected by previous works due to inaccurate degradation cost modeling; 3) BESS regulation revenue depends on charge/discharge limit; 4) BESS reserve revenue depends on charge/discharge limit and capacity; 5) BESS energy revenue depends on charge/discharge limit, capacity, and replacement cost; 6) degradation cost modeling impacts BESS participation strategies in various markets, which is further discussed in \cite{NAPS2}.

\section{Conclusion}

This paper proposes a bi-level framework for optimal strategies of price-maker BESSs in real-time energy, spinning reserve, and frequency regulation markets. To understand BESSs' excessive regulation market activities and the impact of degradation on BESS market strategies, an accurate degradation cost function is deployed in the BESS strategic bidding model, and an AGC signal dispatch model is proposed to model and simulate BESS AGC signal following activities. Case studies on a synthetic system with real-world data are performed to investigate BESS operating characteristics in the regulation market and energy/reserve markets, when BESS degradation, AGC signal following, and system annual load/ancillary service requirements variations are considered. Impacts of BESS capacity and replacement cost on its market revenue are also investigated.

Future research could focus on considering market uncertainties and renewable energy penetration for better evaluation of BESS's impact on energy and ancillary services markets.
\appendix


 \bibliographystyle{elsarticle-num} 
 \bibliography{cas-refs}





\end{document}